\documentstyle[aps,preprint,tighten,floats,epsf,rotate]{revtex}

\begin{document}
\draft

\preprint{\vbox{\it 
                        \null\hfill\rm    IP-BBSR/97-15, April'97}\\\\}
%
\title{Surface Induced Phase Transition in Quark-Gluon Plasma 
Produced in Laboratory}
%
\author{Sanatan Digal and Ajit M. Srivastava}
\address{Institute of Physics, Bhubaneswar 751005, India}
%
%
\maketitle
\widetext
\parshape=1 0.75in 5.5in
\begin{abstract}
We discuss an {\it outside-inside} scenario for a first order 
quark-hadron transition in the Quark-Gluon Plasma (QGP) expected
to be produced in heavy ion collisions, wherein the entire QGP region 
itself becomes like a subcritical bubble, and starts shrinking. We 
argue that this shrinking QGP bubble will lead to concentration 
of baryon number in a narrow beam like region in center, 
which can be detected by HBT analysis,  or as a raised plateau
in the rapidity plot of baryon number.
\end{abstract}
\vskip 0.125 in
\parshape=1 -.75in 5.5in
\narrowtext

 Possibility of observing signatures of a quark-hadron transition
in relativistic heavy-ion collisions has been extensively investigated 
for sometime now. Lattice calculations allow for the possibility that 
this transition may be of  first order. Earlier studies of such a first
order transition have primarily focused on employing homogeneous 
nucleation of hadronic bubbles in a uniform QGP background. These
bubbles expand after nucleation and release latent heat which heats
up the plasma, suppressing further nucleation \cite{kpst}. 

 We discuss an alternative scenario for a first order quark-hadron 
transition in this paper by focusing on the finite size of the QGP 
region produced in heavy ion collisions. Outside this region, one 
either has a hadronic gas (for example, due to energy density 
gradient in the transverse direction \cite{denstr}), or simply the 
QCD vacuum in the confining phase. In either case, there must be a 
domain wall at the boundary of the QGP region with a non-zero surface
tension. For simplicity, we take the region outside this boundary 
to be (at least a thin layer of) the hadronic phase with roughly 
same temperature as inside.

 We solve the hydrodynamical equations for the longitudinal expansion
model of Bjorken \cite{bj}, and show that transition can proceed 
entirely due to collapse of this boundary wall.
We argue that collapsing boundary wall will lead to concentration 
of baryons in the central region in a narrow beam like region (of  
thickness 1-2 fm and length of few tens of fm), which can be detected 
by the HBT analysis \cite{hbt} of baryons, or by a raised plateau in 
the rapidity plot of baryon number.
 
 It is important to appreciate that, in contrast to usual QGP bubbles 
which nucleate by thermal fluctuations in a superheated hadronic matter,
this  entire, bubble like, region of QGP does not arise due to any 
nucleation process. Rather, it forms due to certain initial conditions 
which create a dense, interacting gas of partons in a collision process 
and this gas thermalizes later to form a region of QGP. As thermalization
proceeds, the order parameter develops a value appropriate for
high temperature, leading to the formation of the interface at
the boundary. 

  We start by writing down the hydrodynamical equations governing the 
evolution of energy density $e$ during the longitudinal 
expansion \cite{bj}.

\begin{equation}
\label{evol}
{de \over d\tau} = - {(e + p) \over \tau} .
\end{equation}

 Here $\tau$ denotes the proper time and $p$ is the pressure. [We will 
use natural system of units with $\hbar = c = 1$.] At temperatures 
well above the critical temperature, $T_c$, QGP will keep expanding 
in the longitudinal direction without any hindrance. [We neglect any 
transverse expansion.] As the temperature drops due to the longitudinal 
expansion via eqn.(\ref{evol}), the outward expansion of the wall (in 
the local rest frame) will become energetically unfavorable at some 
temperature greater than $T_c$. Let us first estimate this temperature 
for a simple case when the QGP region is spherical with radius $R$. 
The free energy associated with such a bubble is

\begin{equation}
\label{free}
F = -{4\pi \over 3} [p_q(T) - p_h(T)] R^3 + 4\pi R^2 \sigma ,
\end{equation}

\noindent where $p_q$ and $p_h$ are the pressures in the QGP and hadronic 
phases respectively and $\sigma$ is the surface tension of the 
wall. Taking the hadronic gas to consist of three massless pions and QGP to
consist of gluons and 2 flavors of massless quarks, we have

\begin{equation}
\label{pqph}
p_q(T) = {37\pi^2 \over 90} T^4 - B, \quad p_h(T) = {3\pi^2 \over 90} T^4 ,
\end{equation}

\noindent where $B$ is the bag constant (which is related to $T_c$ by 
Gibb's condition, $p_q(T_c) = p_h(T_c)$). For $T > T_c$, the critical 
radius of the QGP bubble is,

\begin{equation}
\label{Rc}
R_c(T) = {2\sigma \over p_q(T) - p_h(T)} .
\end{equation}

 A QGP bubble of radius larger than $R_c$ will expand while bubbles
of smaller sizes will be subcritical and collapse. By writing $T = 
T_c + \bigtriangleup T$ and assuming $\bigtriangleup T / T_c << 1$ 
(which will be the case of interest to us) we can find the expression for
$\bigtriangleup T$ such that for $T < T_c + \bigtriangleup T$ the radius 
of the QGP region, $R$, becomes smaller than the critical radius $R_c$. 
We find,

\begin{equation}
\label{DTsph}
\bigtriangleup T = {180 \sigma \over 136 \pi^2 R T_c^3} 
\equiv \bigtriangleup T_{sph} .
\end{equation}

We take $\sigma$ = 50 MeV/$fm^2$ \cite{sgma}. With $T_c$ = 170 MeV, we 
find $\bigtriangleup T \simeq 2 MeV$, for a spherical QGP region with
$R$ = 5 fm. So this region starts collapsing when the temperature drops 
below 172 MeV. For $T_c$ = 120 MeV, we get $\bigtriangleup T \simeq$ 
5 MeV. Hereafter, we use $T_c$ = 170 MeV.

  We now estimate the values of $\bigtriangleup T$ suitable for a more 
realistic cylindrical geometry of the QGP region. The free energy of 
this QGP region (with $R_T$ being the transverse radius and
$R_L$ the half length) can be written as 

\begin{equation}
\label{fcyl}
F = - [p_q(T) - p_h(T)] 2 \pi R_T^2 R_L + \sigma
4\pi R_T R_L + 2 \sigma \pi R_T^2 .
\end{equation}

 Here, we have taken the end caps of the cylindrical region to be 
approximated by flat disks of radius $R_T$. As the dynamics of wall 
will be governed by local geometry alone, we only keep relevant 
terms in eqn.(\ref{fcyl}), for a given portion under consideration,
and discard others. We first determine the critical size for the 
middle region, near (longitudinal coordinate ) $z = 0$. For this, 
we neglect the end portion of the cylinder, i.e. the last term in 
eqn.(\ref{fcyl}). The condition for the transverse expansion to be 
energetically favored is,

\begin{equation}
\label{dfdrt}
{\partial F \over \partial R_T} < 0 .
\end{equation}

 This gives a critical value of $R_T$ = $R_c/2$ where $R_c$ is given by 
eqn.(\ref{Rc}). Thus for a given value of $R_T$, when the temperature 
drops below a critical value $T_c + \bigtriangleup T_{tr}$ then the 
transverse dimension becomes subcritical and the wall should start 
collapsing. Using arguments as for eqn.(\ref{DTsph}), we find that 
$\bigtriangleup T_{tr} = \bigtriangleup T_{sph}/2$.

  For the longitudinal motion of the wall, we analyze things in the 
local frame of the plasma. The relevant free energy of 
a small portion of the cylinder at the end region (with length being 
$\bigtriangleup z$, say) is given by terms in eqn.(\ref{fcyl}) with $R_L$ 
replaced by $\bigtriangleup z$. The condition for the longitudinal 
expansion of the wall to be energetically favored is now that the partial
derivative of $F$ with respect to $\bigtriangleup z$ be less than zero
(similar to eqn.(\ref{dfdrt})). This leads to the condition, $R_T > R_c$. 
[Here, it is helpful to think of the end caps to have typical curvature
determined by $R_T$, instead of being strictly flat.] Again, for a given 
value of $R_T$, when the local temperature drops below a critical value 
$T_c + \bigtriangleup T_{long}$ then the longitudinal dimension becomes 
subcritical and the wall starts shrinking in that direction. Here we 
find that $\bigtriangleup T_{long} = \bigtriangleup T_{sph}$. 

 We now come back to eqn.(\ref{evol}) for the evolution of the
energy density. If the total volume of the region (including the
QGP region and the hadronic region) is $V_0$ and the volume of
the QGP region is $V_q$ then we can write the average energy 
density as \cite{kpst},

\begin{equation}
\label{edns}
e(\tau) = e_q {V_q \over V_0} + e_h (1 - {V_q \over V_0}) ,
\end{equation}

\noindent where $e_q$ and $e_h$ are the energy densities in the 
QGP phase and the hadronic phase respectively,

\begin{equation}
e_q = {37 \over 30} \pi^2 T^4 + B , \quad e_h = {\pi^2 \over 10} T^4 .
\end{equation}

 By writing a similar expression, as in eqn.(\ref{edns}), for the 
enthalpy ($e + p$), eqn.(\ref{evol}) leads to the evolution equation for 
the temperature as a function of proper time.  Let us first study the 
region near $z = 0$. For a boost invariant geometry, the same temperature
evolution will be observed at different values of $z$. [Again, for central 
portion, we neglect any possible effects of the end cap region.] We 
consider a cylindrical region of a small length $\bigtriangleup z$ around 
$z = 0$. The outer radius  of the cylinder is $R_T$ and the radius of the 
collapsing wall (bounding the QGP region) is denoted by $R_q$. We then 
have $V_q = \pi R_q^2 \bigtriangleup z$ and $V_0 = \pi R_0^2 \bigtriangleup 
z$.  As we are neglecting the transverse expansion of the plasma,
$R_T$ is time independent while $R_q$ decreases with the velocity of
the wall $v(T)$ moving through plasma. That is, $\dot{R}_q = -v(T)$.
For $v(T)$ we take the following expression

\begin{equation}
\label{vel}
v(T) = a [1 - {T \over T_c}]^b .
\end{equation}

 Here a and b are parameters and $T \le T_c$. There are many uncertainties 
in determining the velocity of the wall such as the energy flux through 
the interface, proper account of surface tension etc.\cite{vwall}. We 
will use two sets of values of these parameters. For one set we use 
$a = 3, b = 3/2$ (valid for $T > 2T_c/3$, see \cite{kpst,vwall}) and 
for the other set we use a sample value, $a$ = 2 and $b = 1$ (this 
should be used for $T >$ about $0.7T_c$, so that $v(T)$ remains less 
than the speed of sound). [Velocities for this second set are larger 
than those used in the literature \cite{kpst}, though linear dependence 
on $T$ has been discussed in \cite{vkj}. We use this set as an example,
since we expect that the collapse of the wall will be faster with 
respect to local plasma when one allows for even slightest expansion 
of the plasma with respect to wall, especially near $T = T_c$.]

 There is one subtle point here. The velocity as given by 
eqn.(\ref{vel}) vanishes for $T = T_c$. However, we have argued above 
that a QGP region of finite size becomes subcritical at a temperature  
{\it larger} than $T_c$. Therefore, in this case, the velocity 
of the wall (with respect to the local plasma) should be more
accurately  represented when we replace $T_c$ in 
eqn.(\ref{vel}) by $T_c + \bigtriangleup T$. That is,

\begin{equation}
\label{veldt}
v(T) = a[1 - {T \over T_c + \bigtriangleup T}]^b ,
\end{equation}

\noindent valid for $T \le T_c + \bigtriangleup T$, here, 
$\bigtriangleup T = \bigtriangleup T_{tr}$, or 
$\bigtriangleup T_{long}$, depending on whether one is discussing the 
wall motion in the transverse direction or in the longitudinal direction.
[We mention here that the qualitative aspects of
our results do not depend on whether one uses eqn.(\ref{vel}) or
eqn.(\ref{veldt}) for wall velocity.] 
    
   Figure 1 shows the evolution of the temperature near $z = 0$ region.
Initial conditions, prescribed at (a sample value of) $\tau_0$ = 3 fm, 
are, $T(\tau_0) = T_c + \bigtriangleup T$, and $R_q(\tau_0) = R_T$. The 
plots are given only up to those specific values of proper time, when 
the transverse dimension of the QGP region, $R_q$, becomes zero. 
   
 For the evolution of temperature near the
end regions of the plasma in the longitudinal direction, 
consider a segment of length $z_0(\tau)$ at the end portion 
with the transverse dimension of the region being $R_T$. 
The location of the end portion of the longitudinally moving 
wall is denoted by $z_q(\tau)$ while the transverse radius of the
wall is given by $R_q(\tau)$.  The ratio of the volume of the 
QGP region $V_q$ to the full (local) region $V_0$ is then given by

\begin{equation}
\label{endp}
{V_q \over V_0}  = {\pi R_q(\tau)^2 z_q(\tau) \over 
\pi R_T^2 z_0(\tau)} . 
\end{equation}

 Here, $z_0(\tau)$ increases due to plasma expansion with the
velocity $\dot{z}_0 = z_0/{\tau}$. Wall velocity in local rest frame
of the plasma is again given by eqn.(\ref{veldt}), so $\dot{R}_q
= - v(T)$ and $\dot{z}_q = z_q/{\tau} - v(T)$ (as both velocities are
small).  For simplicity, we use a single value of $\bigtriangleup T$ 
($= \bigtriangleup T_{long}$) for both directions.
With the same initial conditions, as in Fig.1, we then determine
the evolution of temperature in the end region. We find, that the 
evolution of temperature is almost similar here as for the $z = 0$ 
region (Fig.1), apart from minor differences, (thus we do not show
these plots). For example, corresponding to the dotted curve in 
Fig.1, one finds little larger re-heating for the end portion, 
while for the thick solid curve of Fig.1, one finds smaller 
reheating. [Note that the temperature 
obtained from the evolution equation represents an average 
temperature for the cylindrical segment, and hence, is valid 
only for time scales larger than the one given by $R_T$.]

\begin{figure}[h]
\begin{center}
\vskip -1 in
\leavevmode
\epsfysize=10truecm \vbox{\epsfbox{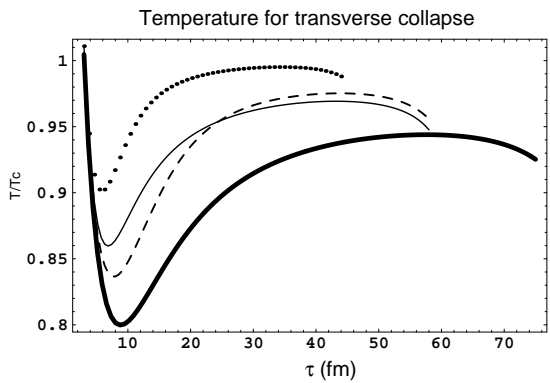}}
\vskip -1in
\end{center}
\caption{Temperature in the central region as a function of proper 
time. In both of the figures in this paper, thick solid curve 
and dashed curve correspond to $R_T$ = 7 fm, while thin solid curve 
and dotted curve correspond to $R_T$ = 3 fm. Solid curves 
correspond to the choice of parameters for the wall velocity 
(eqn.(\ref{veldt})) as, $a = 3, b = 3/2$, while dotted and dashed 
curves correspond to the parameter set $a = 2, b = 1$.}
\label{Fig.1} 
\end{figure}
    
   We see from Fig.1 that largest reheating temperature is represented by 
the dotted curve. First, $T$ drops sharply 
to about 0.9$T_c$ due to expansion of 
the plasma. This value is low enough to cause nucleation of bubbles in 
the central region, see \cite{kpst}. Note that the heating, and hence 
suppression of bubble nucleation, will be somewhat larger near the wall as 
compared to the center. [We assume that reheating is dominated by the 
collapsing boundary wall due to large surface area, compared to any nucleated 
bubbles.] For $\tau >$ about 6 fm, the reheating dominates over cooling 
due to the expansion, and $T$ starts to rise. It reaches a value of 
about 0.98$T_c$ by $\tau \simeq$ 16 fm and 0.99$T_c$ by $\tau \simeq$ 
21 fm. Critical bubble sizes corresponding to these temperatures are, 
$R_c \simeq 3.4$ fm and 6.7 fm, respectively (eqn.(\ref{Rc})). The entire 
size of the region is not very large, or, more importantly,  the elapsed 
time is not enough to let any bubbles, nucleated earlier in the interior 
(when the temperature was lower there), to grow to these sizes. [Note 
that the wall velocity of the bubbles, eqn.(\ref{vel}), (or 
eqn.(\ref{veldt}) with appropriate $\bigtriangleup T$), is very small
for temperatures close to $T_c$.] Thus, all these bubbles become 
subcritical and collapse away. The end result being that the phase 
transition to the hadronic phase proceeds entirely due to the collapse 
of the outer wall.

 For some other parameters, e.g. those corresponding to the plot given 
by thick solid curve in Fig.1, the final reheat temperature 
is not large enough to completely quench nucleation. In this case, 
the picture of the transition may be somewhat mixed. 

  Let us now discuss about some possible observable consequences of
this type of scenario of phase transition. Here we apply ideas used 
in the early Universe in the context of baryon number inhomogeneity
generation at the end of quark-hadron transition \cite{unvrs}. Due
to large masses of baryons in the hadronic phase and much smaller
masses of quarks in the QGP phase, the baryon number density in
the QGP phase is larger than in the hadronic phase. The ratio
$R = n^B_q/n^B_h$ of the baryon number densities in the two phases 
has been estimated by many people. A rough estimate may give
$R \simeq 10$.  Evolution of the value of $R$ in a collapsing QGP 
droplet has also been extensively discussed in literature in the 
context of the early Universe \cite{unvrs}.  It seems reasonable that a 
good fraction of the net baryon number inside the droplet may get trapped 
inside. We will simply assume this to be the case, at this preliminary 
stage of exploration, instead of getting into details of the mechanism of 
trapping. 

 In our model of the phase transition in a heavy ion collision, the
shrinking of boundary wall leads to a picture which is identical to
that of a shrinking QGP droplet in the early Universe. Immediate
consequence is that the net baryon number of the region (which will 
be expected to be non-zero, though small, even for RHIC and LHC
energies) will become concentrated near the central region 
of the plasma due to the collapse of the boundary wall. 
[We mention here that people have discussed formation of
strangelets due to strangeness enhancement of the QGP droplet in
heavy ion collisions by evaporating hadrons \cite{strng}. However,
as we show below, the longitudinal collapse of the wall is insignificant 
by the time transition ends, i.e. when transverse collapse of wall is
completed. Thus, the picture of any, strangeness enriched, exotic
lump of matter forming at the center of the collision, or in
different portions due to local fluctuations, needs to be modified.]  

  We calculate the longitudinal collapse of the wall, in the center of
mass frame, by taking its velocity with respect to local plasma to be 
given by eqn.(\ref{veldt}).  In the longitudinal expansion model, the 
plasma velocity at longitudinal distance $z$ is given by $z/t$ \cite{bj}. 
Taking the end portion of the wall to have initial rapidity equal to 
some maximum rapidity, the evolution of frame rapidity at wall 
end is shown in Fig.2, 
until the time when transverse collapse is completed. The final value of
this frame rapidity gives the rapidity interval inside which most 
of the net baryon number may be expected to be concentrated.  
We see that this interval can shrink by up to one unit by the 
time the transition is completed, i.e. when $R_q$ becomes zero.

\begin{figure}[h]
\begin{center}
\vskip -1in
\leavevmode
\epsfysize=10truecm \vbox{\epsfbox{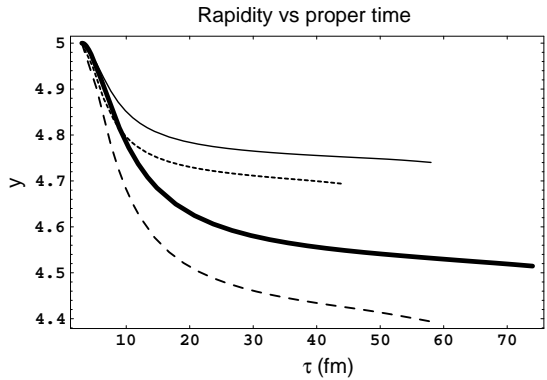}}
\vskip -1in
\end{center}
\caption{Evolution of frame rapidity at the end portion of the wall 
due to its longitudinal collapse in the center of mass frame.}
\label{Fig.2} 
\end{figure}

 Large values of the final rapidity of the wall imply that the 
longitudinal extent of the wall is large when the transverse collapse 
is completed. This concentrates all (or most) of the baryon number in a 
narrow beam like structure at the center of the cylindrical region. 
Transverse dimension of this region should be of the order of wall 
thickness, i.e. 1-2 fm, while its length can be very large, of the order 
of $\tau$ when $R_q$ becomes zero in Figs.1 and 2. Such a structure for 
baryon source can be easily detected by the HBT analysis. 

  Further, the reduction in the frame rapidity at the end portion 
of wall implies that in the rapidity plot of 
net baryon number, one will expect a raised plateau near zero 
rapidity with width determined by the final rapidity in Fig.2. 
Equivalently, one may expect a depression in the rapidity plot of 
antibaryons due to annihilations with increased number of baryons.
Uniformity of this plateau will depend on whether some (or all) of 
the fragmentation region lies inside the boundary wall initially. 

 We conclude by summarizing our main results. We have presented
a novel scenario of first order quark-hadron phase transition 
relevant for the situation in heavy ion 
collisions. Here, the phase transition proceeds by the shrinking 
of the interface surrounding the entire QGP region, starting 
at a temperature slightly above $T_c$ itself (due to the QGP region
becoming subcritical). Collapsing boundary wall heats up the 
entire region due to the release of latent heat which inhibits 
bubble nucleation in the enclosed region. The reheat temperature 
can be close enough to $T_c$ so that if any bubbles were nucleated
in the central region, they become subcritical and shrink away. Thus, 
finally, the entire phase transition proceeds due to collapse of the 
boundary wall only. We argue that collapsing  boundary wall can 
concentrate the net baryon number of the QGP region in center in a 
long beam like region of very small thickness (of order 1-2 fm), which 
can be detected by HBT analysis. Also, as baryons get concentrated in 
a smaller rapidity interval, it should be observable as a raised plateau 
in the rapidity plot of baryon number (or a depressed plateau 
in the rapidity plot of antibaryons).  We emphasize that a signature of
a pencil shaped source of baryon number will not only demonstrate 
the existence of QGP state, it will also show the transition to be of
first order, and will provide experimental evidence for the baryon 
concentration by shrinking QGP bubbles as has been proposed for the
early Universe. 

 Other possible signatures of this picture of phase transition can arise 
in rapidity plots of strange particles, in transverse momentum distributions 
of particles (e.g. baryons) reflecting concentration in the transverse 
direction, etc. It will be interesting to see if the collapsing boundary 
wall can leave some imprints also on the spectrum of photons and dileptons.


 We  would like to thank Pankaj Agrawal, Shashi Phatak, and 
Supratim Sengupta for many useful comments. We are very grateful
to Larry McLerran for many useful suggestions, especially about
the HBT analysis to detect the transverse collapse.

\end{document}